\documentclass[pra,a4paper,oneside,twocolumn,superscriptaddress,showpacs]{revtex4-1}

\usepackage[english]{babel}
\usepackage[latin1]{inputenc}

\usepackage{dcolumn}
\usepackage{subfigure}
\usepackage{multirow}
\usepackage{amsmath,amsfonts,amssymb}

\usepackage{graphicx}  % standard LaTeX graphics tool
\usepackage{latexsym}   % useful for coding complex math
\usepackage{color}

\begin{document}

\title{Unconventional Fano effect and off-resonance field enhancement in plasmonic coated spheres}

\author{\firstname{Tiago}  J. \surname{Arruda}}
\email{tiagoarruda@pg.ffclrp.usp.br}
\affiliation{Faculdade de Filosofia,~Ci\^encias e Letras de Ribeir\~ao
Preto, Universidade de S\~ao Paulo, 14040-901, Ribeir\~ao
Preto-SP, Brazil}

\author{\firstname{Alexandre} S. \surname{Martinez}}
%\email{asmartinez@ffclrp.usp.br}
\affiliation{Faculdade de Filosofia,~Ci\^encias e Letras de Ribeir\~ao
Preto, Universidade de S\~ao Paulo, 14040-901, Ribeir\~ao
Preto-SP, Brazil}
\affiliation{National Institute of Science and Technology in
Complex Systems}

\author{\firstname{Felipe}  A. \surname{Pinheiro}}
\affiliation{Instituto de F\'{i}sica, Universidade Federal do Rio de Janeiro, 21941-972, Rio de Janeiro-RJ, Brazil}

\begin{abstract}
We investigate light scattering by coated spheres composed of a dispersive plasmonic core and a dielectric shell.
By writing the absorption cross-section in terms of the internal electromagnetic fields, we demonstrate it is an observable sensitive to interferences that ultimately lead to the Fano effect.
Specially, we show that unconventional Fano resonances, recently discovered for homogeneous spheres with large dielectric permittivities, can also occur for metallic spheres coated with single dielectric layers.
These resonances arise from the interference between two electromagnetic modes with the same multipole moment inside the shell and not from interactions between various plasmon modes of different layers of the particle.
In contrast to the case of homogeneous spheres, unconventional Fano resonances in coated spheres exist even in the Rayleigh limit.
These resonances can induce an off-resonance field enhancement, which is approximately one order of magnitude larger than the one achieved with conventional Fano resonances.
We find that unconventional and conventional Fano resonances can occur at the same input frequency provided the dispersive core has a negative refraction index.
This leads to an optimal field enhancement inside the particle, a result that could be useful for potential applications in plasmonics.

\end{abstract}


%\pacs{290.0290, % Scattering
%      290.4020, % Mie theory
%      290.5825, % Scattering theory
%      290.5850. % scattering, particles
%     }

\maketitle

%\tableofcontents

\section{Introduction}

The Fano resonance, discovered in the realm of atomic physics by U. Fano in 1961~\cite{fano1961}, is one distinctive characteristic of interacting quantum systems.
The unique, asymmetric Fano lineshape has its origin in the wave interferences between a narrow discrete resonance and the continuum or a broad resonance.
As a signature of quantum interference, the Fano effect has been extensively investigated in electronic transport at the nanoscale, in systems such as quantum dots, quantum wires, and tunnel junctions (for a review see Ref.~\cite{mirormp}).

Being an interference phenomenon, Fano resonances also manifest themselves in classical optics.
Historically, the first observation of a Fano resonance in optics was probably the Wood's anomaly~\cite{wood}.
With the advent of metamaterials and plasmonic nanostructures, the Fano effect has recently become an important tool to control electromagnetic mode interactions~\cite{luk}.
Due to the sharpness of the Fano asymmetric lineshape, systems exhibiting the Fano effect are highly sensitive to the local dielectric environment.
As a consequence, in plasmonic systems the Fano effect has been exploited in the development of optical sensors, nonlinear devices, and low-threshold nanoscopic lasers~\cite{luk}.
Fano resonances have also been observed in several photonic systems, such as micropillars, photonic crystals, sub wavelength apertures in polaritonic membranes, metallic films, and random photonic structures (see Ref.~\cite{luk} and references therein).

In light scattering by small particles, Fano resonances are also known to play an important role.
Despite its long history since the pioneer work of Mie in 1908~\cite{bohren}, light scattering still reveals challenging surprises, and many among them are related to the Fano effect.
Indeed, anomalous light scattering phenomena such as giant optical resonances with an inverse hierarchy (quadrupole resonance is stronger than the dipole one), near-field pattern exhibiting vortices, and unusual size and frequency dependencies~\cite{Tribelsky2006,bashevoy2005,zhang2004} were found to be described in terms of an analogy with Fano resonances of a quantum particle scattered by a potential with quasi-bound levels~\cite{tribelsky2008}.
According to this analogy, localized plasmons (polaritons) excited by an incident radiation play the role of quasi-bound levels and their radiative decay is equivalent to the tunneling process from such levels~\cite{tribelsky2008}.
The interference between the incident and re-emitted radiation gives rise to either enhancement (constructive interference) or suppression (destructive interference) of the electromagnetic field.
Fano resonances were shown to be at the origin of the so-called ``dark-states'' in light scattering by coated spheres, where the local electromagnetic field enhancement occurs in off-resonant regions~\cite{miroshnichenko}.
Light scattering by homogeneous particles can also exhibit unconventional Fano resonances in the extinction cross-section beyond the Rayleigh approximation; they result from the interference between different electromagnetic modes with the same multipole moment~\cite{tribelsky}.
This contrasts to conventional Fano resonances, which arise from the spectral overlap of broad and narrow electromagnetic modes with different values of multipole moment~\cite{tribelsky}.

The aim of this paper is to investigate conventional and unconventional Fano resonances and their connection to off-resonance field enhancement in coated spheres composed of a plasmonic core and a dielectric shell.
We demonstrate that unconventional Fano resonances can also show up for coated spheres even in the Rayleigh limit, which is in contrast to the case of homogeneous spheres.
These unconventional Fano resonances are at the origin of an off-resonance field enhancement within the scatterer, which can be even larger than the one achieved with conventional Fano resonances.
In addition, we show that when the core has negative refractive index, conventional and unconventional Fano resonances can occur at the same input frequency, corresponding to the condition for an optimal field enhancement within the particle.

This paper is organized as follows.
In Sec.~\ref{aden-kerker}, we describe light scattering by coated magnetic spheres within a generalization of the Lorenz-Mie scattering.
In Sec.~\ref{fano-sec}, we investigate the behavior of conventional and unconventional Fano resonances in coated spheres.
The connection to off-resonance field enhancement is discussed in Sec.~\ref{off-res}, whereas Sec.~\ref{negative} is devoted to the case of cores with negative refractive index.
Finally, in Sec.~\ref{conclusions}, we summarize our main results and conclude.

\section{Light scattering by a coated sphere}
\label{aden-kerker}

\begin{figure}[htb!]
\includegraphics[angle=0, width=.7\linewidth]{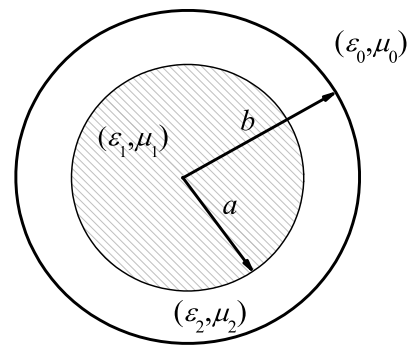}
\caption{The scatterer geometry: a sphere of radius $a$ and optical properties $(\epsilon_1,\mu_1)$ coated with a spherical shell of thickness $(b-a)$ and optical properties $(\epsilon_2,\mu_2)$; the embedding medium is $(\epsilon_0,\mu_0)$.}
\label{fig1}
\end{figure}

Let us consider the electromagnetic wave scattering by a coated sphere within the framework of the Aden-Kerker generalization of the Lorenz-Mie solution~\cite{aden,bohren}.
This theory describes the scattering of a plane electromagnetic wave $(\mathbf{E},\mathbf{H})e^{-\imath\omega t}$, with $\omega$ the angular frequency, by a coated sphere with inner radius $a$ and outer one $b$~\cite{aden,bohren}.
We assume that the coated sphere is made of linear, spatially homogeneous, and isotropic materials.
The electric permittivities and magnetic permeabilities of the core ($0\leq r\leq a$) and the shell ($a\leq r\leq b$) are $(\epsilon_1,\mu_1)$ and $(\epsilon_2,\mu_2)$, respectively.
For simplicity, the scatterer is embedded in vacuum, with optical properties $(\epsilon_0,\mu_0)$ [see Fig.~\ref{fig1}].
The extinction and scattering efficiencies (which are the respective cross-sections in units of $\pi b^2$) are: $Q_{\rm ext}=({2}/{y^2})\sum_{n=1}^{\infty}(2n+1){\rm Re}(a_n+b_n)$ and $Q_{\rm sca}=({2}/{y^2})\sum_{n=1}^{\infty}(2n+1)(|a_n|^2+|b_n|^2)$, where $y=kb$ is the size parameter of the outer sphere ($k$ being the incident wave number)~\cite{bohren}.
By energy conservation, the absorption efficiency is $Q_{\rm abs}=Q_{\rm ext}-Q_{\rm sca}$.
In particular, the radar backscattering efficiency is $Q_{\rm back}=({1}/{y^2})|\sum_{n=1}^{\infty}(-1)^n(2n+1)(a_n-b_n)|^2$.
In all the $Q$ factors, the Aden-Kerker scattering coefficients $a_n$ and $b_n$ are~\cite{tiago-joa}:
\begin{eqnarray}
    a_n&=&\frac{\left(\widetilde{D}_n/\widetilde{m}_2+n/y\right)\psi_n(y)-\psi_{n-1}(y)}{\left(\widetilde{D}_n/\widetilde{m}_2+n/y\right)\xi_n(y)-\xi_{n-1}(y)}\
    ,\label{an}\\
    b_n&=&\frac{\left(\widetilde{m}_2\widetilde{G}_n+n/y\right)\psi_n(y)-\psi_{n-1}(y)}{\left(\widetilde{m}_2\widetilde{G}_n+n/y\right)\xi_n(y)-\xi_{n-1}(y)}\ ,
\end{eqnarray}
where one defines the auxiliary functions
\begin{eqnarray}
    \widetilde{D}_n&=&\frac{D_n(m_2y)-A_n\chi_n'(m_2y)/\psi_n(m_2y)}{1-A_n\chi_n(m_2y)/\psi_n(m_2y)}\
    ,\label{Dn}\\
    \widetilde{G}_n&=&\frac{D_n(m_2y)-B_n\chi_n'(m_2y)/\psi_n(m_2y)}{1-B_n\chi_n(m_2y)/\psi_n(m_2y)}\
    ,\\
    A_n&=&\frac{\psi_n(m_2x)\left[\widetilde{m}_2D_n(m_1x)-\widetilde{m}_1D_n(m_2x)\right]}{\widetilde{m}_2D_n(m_1x)\chi_n(m_2x)-\widetilde{m}_1\chi_n'(m_2x)}\
    ,\label{An}\\
    B_n&=&\frac{\psi_n(m_2x)\left[\widetilde{m}_2D_n(m_2x)-\widetilde{m}_1D_n(m_1x)\right]}{\widetilde{m}_2\chi_n'(m_2x)-\widetilde{m}_1D_n(m_1x)\chi_n(m_2x)}
    ,\label{Bn}
\end{eqnarray}
with $x=ka$ the size parameter of the inner sphere and $D_n(\rho) \equiv {\rm d}[\ln \psi_n(\rho)]/{\rm d}\rho$.
The functions $\psi_n(\rho)=\rho j_n(\rho)$, $\chi_n(\rho)=-\rho y_n(\rho)$ and $\xi_n(\rho)=\psi_n(\rho)-\imath\chi_n(\rho)$ are the Riccati-Bessel, Riccati-Neumann and Riccati-Hankel functions, respectively, with $j_n$ and $y_n$ being the spherical Bessel and Neumann functions.
The refractive and impedance indices are $m_q=k_q/k=\sqrt{\epsilon_q\mu_q/(\epsilon_0\mu_0)}$ and $\widetilde{m}_q=\mu_0 m_q/\mu_q=\sqrt{\epsilon_q\mu_0/(\epsilon_0\mu_q)}$, respectively~\cite{fap_prl} (with $q=1$ for the core and $q=2$ for the shell).

The angle-averaged electric $\langle |\mathbf{E}|^2\rangle_{\rm \Omega}$ and magnetic $\langle |\mathbf{H}|^2\rangle_{\rm \Omega}$ fields within the coated sphere can be explicitly calculated~\cite{kaiser}.
Here, we define $\langle |\mathbf{F}|^2\rangle_{\rm \Omega}\equiv \int_0^{2\pi} {\rm d}\phi \int_{-1}^{1} {\rm d}(\cos\theta)\ |\mathbf{F}|^2/4\pi$,
where $\mathbf{F}=\mathbf{F}(r,\cos\theta,\phi)$ is either the electric ($\mathbf{E}$) or the magnetic ($\mathbf{H}$) fields in the spherical coordinate system~$(r,\theta,\phi)$.
Using the exact expression for the electric field inside the core region $(0\leq r\leq a)$, we obtain~\cite{tiago-sphere,tiago-joa}:
\begin{equation}\begin{split}
    \frac{\langle |\mathbf{E}_1|^2\rangle_{\rm \Omega}}{|E_0|^2}=&\frac{1}{2}\sum_{n=1}^{\infty}\bigg\{(2n+1) |c_n |^2 |j_n(\rho_1)|^2 \\
    &+ |d_n |^2\left[n
    |j_{n+1}(\rho_1)|^2+(n+1)|j_{n-1}(\rho_1)|^2\right]\bigg\}\ ,\label{E1-med}
\end{split}\end{equation}
where $\rho_1=m_1kr$ and $E_0$ is the incident wave amplitude.
The expression for the magnetic field is obtained by replacing $(\mathbf{E}_1,E_0)$ with $(\mathbf{H}_1,H_0)$ (where $H_0=E_0\sqrt{\epsilon_0/\mu_0}$), and $(c_n,d_n)$ with $(\widetilde{m}_1 d_n,\widetilde{m}_1 c_n)$ in Eq.~(\ref{E1-med}).
For the shell region ($a\leq r\leq b$) the angle-averaged electric field is~\cite{tiago-joa}:
\begin{eqnarray}
    \frac{\langle |\mathbf{E}_2|^2\rangle_{\rm \Omega}}{|E_0|^2}&=&\frac{1}{2}\sum_{n=1}^{\infty}\Bigg\{(2n+1) \left[|f_n |^2 |j_n(\rho_2)|^2+  |v_n |^2 |y_n(\rho_2)|^2\right]\nonumber\\
    &&+ |g_n |^2 \left[n
    |j_{n+1}(\rho_2)|^2+(n+1)|j_{n-1}(\rho_2)|^2
    \right]\nonumber\\
    &&+ |w_n |^2 \left[n
    |y_{n+1}(\rho_2)|^2+(n+1)|y_{n-1}(\rho_2)|^2
    \right]\nonumber\\
    &&+ 2{\rm Re}\bigg[ (2n+1) f_n v_n^{*} j_n(\rho_2)y_n(\rho_2^*)\nonumber\\
    &&+ g_n w_n^{*}\big[n j_{n+1}(\rho_2)y_{n+1}(\rho_2^*)\nonumber\\
    &&+(n+1)j_{n-1}(\rho_2)y_{n-1}(\rho_2^*)
    \big]\bigg]\Bigg\}\ ,\label{E2-med}
\end{eqnarray}
where $\rho_2=m_2kr$.
The expression for the angle-averaged magnetic field within the shell is obtained by replacing $(\mathbf{E}_2,E_0)$ with $(\mathbf{H}_2,H_0)$, $(f_n,g_n)$ with $(\widetilde{m}_2 g_n,\widetilde{m}_2 f_n)$, and $(v_n,w_n)$ with $(\widetilde{m}_2 w_n,\widetilde{m}_2 v_n)$ in Eq.~(\ref{E2-med}).
In terms of the auxiliary functions defined in Eqs.~(\ref{Dn})--(\ref{Bn}), the internal coefficients $c_n$, $d_n$, $f_n$, $g_n$, $v_n$ and $w_n$ read~\cite{tiago-joa}
    \begin{eqnarray}
        c_n &=&{m_1f_n\left[\psi_n(m_2y)-B_n\chi_n(m_2y)\right]}\left[{m_2\psi_n(m_1x)}\right]^{-1}\
        ,\\
        d_n &=&{m_1g_n\left[\psi_n'(m_2y)-A_n\chi_n'(m_2y)\right]}\left[{m_2\psi_n'(m_1x)}\right]^{-1}\
        ,\\
    f_n&=&\frac{\imath
    m_2/\left[B_n\chi_n(m_2y)-\psi_n(m_2y)\right]}{\left(\widetilde{m}_2\widetilde{G}_n+n/y\right)\xi_n(y)-\xi_{n-1}(y)}\
    ,\\
    g_n&=&\frac{\imath
    m_2/\left[A_n\chi_n(m_2y)-\psi_n(m_2y)\right]}{\left(\widetilde{D}_n+n\widetilde{m}_2/y\right)\xi_n(y)-\widetilde{m}_2\xi_{n-1}(y)}\
    ,\label{gn} \\
        v_n &=&B_nf_n \ ,\\
        w_n &=&A_ng_n \ .
    \end{eqnarray}
The above expressions for the angle-averaged fields inside a coated sphere agree with Ref.~\cite{kaiser} in the nonmagnetic case ($\mu_1=\mu_2=\mu_0$).

\section{Fano-like resonances in coated spheres}
\label{fano-sec}

The identification of conventional Fano resonances require observables that are sensitive to wave interference.
As a result, in light scattering by single homogeneous spheres, Fano-like resonances do not occur in the extinction $Q_{\rm ext}$ and scattering $Q_{\rm sca}$ efficiencies (and consequently in the absorption efficiency, $Q_{\rm abs}$) since these quantities are proportional to the sum of intensities~\cite{luk}.
Also, since both $Q_{\rm ext}$ and $Q_{\rm sca}$ are averaged among all possible directions and polarizations, the resonance properties are averaged as well.
For this reason, conventional Fano resonances in homogeneous spheres are expected to occur only in the cross-sections at a particular direction, {\it e.g.} the backscattering cross-section.
However, for coated spheres the wave interferences, which ultimately lead to Fano-like resonances, can show up in the absorption efficiency $Q_{\rm abs}$ even though $Q_{\rm ext}$ and $Q_{\rm sca}$ do not explicitly contain interference terms.
To demonstrate this, note that one can explicitly write $Q_{\rm abs}$ in terms of internal electromagnetic modes~\cite{tiago-joa,tiago-sphere,ruppin-energy}.
Defining the volume $V_q\equiv 4\pi(l_2^3-l_1^3)/3$, where $(l_1,l_2)=(0,a)$ for the core~($q=1$) and $(l_1,l_2)=(a,b)$ for the shell~($q=2$), one calculates the volume-averaged field intensities using
$\langle |\mathbf{F}|^2\rangle_{V_q} \equiv {3}\int_{l_1}^{l_2}{\rm d}r\ r^2\langle|\mathbf{F}|^2\rangle_{\Omega}/{(l_2^3-l_1^3)}$~\cite{tiago-joa}.
Since $\epsilon_q=\epsilon_q'+\imath\epsilon_q''$ and $\mu_q=\mu_q'+\imath\mu_q''$ and using the relations between $Q_{\rm abs}$ and the absorbed powers~\cite{ruppin-energy},
$P_{{E}_q}=\omega\epsilon_q''\int_{V_q}{\rm d}^3r|\mathbf{E}_q|^2/2$ and $P_{{H_q}}=\omega\mu_q''\int_{V_q}{\rm d}^3r|\mathbf{H}_q|^2/2$, we obtain:
\begin{equation}
\begin{split}
    Q_{\rm
    abs}=&\frac{4}{3}y\bigg[\left(\frac{\epsilon_1''}{\epsilon_0}\frac{\langle|\mathbf{E}_1|^2\rangle_{V_1}}{|E_0|^2}+\frac{\mu_1''}{\mu_0}\frac{\langle|\mathbf{H}_1|^2\rangle_{V_1}}{|H_0|^2}\right)S^3\\
    &+\left(\frac{\epsilon_2''}{\epsilon_0}\frac{\langle|\mathbf{E}_2|^2\rangle_{V_2}}{|E_0|^2}+\frac{\mu_2''}{\mu_0}\frac{\langle|\mathbf{H}_2|^2\rangle_{V_2}}{|H_0|^2}\right)\left(1-S^3\right)\bigg], \label{Qabs-exact}
\end{split}
\end{equation}
with $S \equiv a/b$ being the thickness ratio.
Inspecting the expression for the internal field [Eq.~(\ref{E2-med})], it is clear that Eq.~(\ref{Qabs-exact}) explicitly contains cross-terms such as $f_nv_n^*$ and $g_nw_n^*$ that physically represent interference between internal modes within the shell.
This result suggests that Fano-like resonances could be detected in $Q_{\rm abs}$ as well for coated spheres.

For single homogeneous spheres, unconventional Fano resonances result from the interference between different electromagnetic modes with the same multipole moment so that they show up in $Q_{\rm ext}$~\cite{tribelsky}.
These Fano-like resonances are expected to exist either in homogeneous spheres with large dielectric permittivity, {\it i.e.} beyond the Rayleigh approximation, or in particles with spatial dispersion~\cite{tribelsky}.
In the former case, such resonances may occur due the interference of {\it e.g.} two dipole modes, the resonant one (excited at a large dielectric permittivity) and the off-resonant Rayleigh one.
The unconventional Fano resonances behavior in coated spheres, to the best of our knowledge, has been not reported so far.

To establish the conditions for the occurrence of Fano-like resonances in coated spheres, let us examine resonances in the multipolar moments $a_n$ and $b_n$, which are related to the surface modes or localized plasmon (polariton) resonances (LPRs).
In Lorenz-Mie scattering, these LPRs are usually associated with a strong enhancement of the electromagnetic field intensity near the scatterer surface.
Analytically, they are determined when denominators of the scattering coefficients $a_n$ and $b_n$ vanish~\cite{bohren}.
For nonmagnetic $(\epsilon_1,\mu_1=\mu_0)$ homogeneous spheres of radius $R$ in the Rayleigh limit ($kR\ll1$ and $|m_1|kR\ll1$), the electric surface modes are $\epsilon_1'=-(n+1)\epsilon_0/n$, with $n$ being the mode number.
The lowest-order mode ($n=1$) corresponds to the mode of uniform polarization throughout the sphere.
The frequency in which it occurs, $\epsilon_1'(\omega_{\rm F})=-2\epsilon_0$, is often referred to as Fröhlich frequency~\cite{bohren}.
For coated spheres in the Rayleigh limit ($y\ll1$, $|m_1|x\ll1$ and $|m_2|y\ll1$), the application of the Fröhlich condition for the first surface mode ($n=1$) results in the following dipole scattering coefficient:
\begin{equation}
\begin{split}
    a_1&\approx  \frac{2\imath y^3}{3}\left[\frac{\left(\epsilon_0-\epsilon_2\right)\left(\epsilon_1+2\epsilon_2\right)-\left(\epsilon_0+2\epsilon_2\right)\left(\epsilon_1-\epsilon_2\right)S^3}
                                          {\left(2\epsilon_0+\epsilon_2\right)\left(\epsilon_1+2\epsilon_2\right)-2\left(\epsilon_0-\epsilon_2\right)\left(\epsilon_1-\epsilon_2\right)S^3}\right] \\
        &=-\frac{2\imath y^3}{3}\left[\frac{\epsilon_2/\epsilon_0-(1-S^3)/(1+2S^3)}{\epsilon_2/\epsilon_0+2(1-S^3)/(1+2S^3)}\right]\left(\frac{\epsilon_1-\epsilon_1^{{\rm(ant)}}}{\epsilon_1-\epsilon_1^{{\rm(res)}}}\right), \label{a1aux}
\end{split}
\end{equation}
where we have kept up to $\mathcal{O}(y^5)$.
In Eq.~(\ref{a1aux}), $\epsilon_1^{\rm(ant)}$ and $\epsilon_1^{\rm(res)}$ are the $\epsilon_1$ values which make the numerator (antiresonance) or the denominator (resonance) of $a_1$ to vanish, respectively.
Taking into account that only the core is dispersive [$\epsilon_1=\epsilon_1(\omega)$] in our geometry, the Fröhlich mode is determined for $\epsilon_1(\omega_{\rm F})=\epsilon_1^{\rm(res)}$.
Assuming that the media are weakly absorbing ($\epsilon_q''\ll\epsilon_q'$), we obtain
\begin{eqnarray*}
\left|a_1(\omega)\right|^2\propto\left[\frac{\epsilon_1(\omega)-\epsilon_1^{\rm(ant)}}{\epsilon_1(\omega)-\epsilon_1^{\rm(res)}}\right]^2 = \frac{\left[X(\epsilon_1)+\eta(\epsilon_1)\right]^2}{X(\epsilon_1)^2+1},\\
\end{eqnarray*}
where $X(\epsilon_1)\equiv({\epsilon_1-\widetilde{\omega}_0})/{\sqrt{\widetilde{\gamma}^2-2\widetilde{\gamma}(\epsilon_1-\widetilde{\omega}_0)}}$ and
$\eta(\epsilon_1)\equiv{\widetilde{\gamma}}/{\sqrt{\widetilde{\gamma}^2-2\widetilde{\gamma}(\epsilon_1-\widetilde{\omega}_0)}}$, with $\widetilde{\omega}_0=[{\epsilon_1^{{\rm(res)}}+\epsilon_1^{{\rm(ant)}}}]/{2}$ and $\widetilde{\gamma}=[{\epsilon_1^{{\rm(res)}}-\epsilon_1^{{\rm(ant)}}}]/{2}$.
A small mismatch between the resonance and antiresonance parameters, $|\widetilde{\gamma}/\epsilon_0|\ll1$ (achieved for $S\ll1$), implies that $X(\epsilon_1)\approx(\epsilon_1-\widetilde{\omega}_0)/|\widetilde{\gamma}|$ and $\eta\approx\widetilde{\gamma}/|\widetilde{\gamma}|$, for $\epsilon_1(\omega)\approx\widetilde{\omega}_0$.
In this situation, $|a_1|^2$, as a function of $\epsilon_1(\omega)$, is a Fano-like resonance, characterized by the following (normalized) lineshape~\cite{luk}:
\begin{eqnarray}
\mathcal{F}_{\rm \eta}(X)=\frac{1}{(1+\eta^2)}\frac{(X+\eta)^2}{(X^2+1)}\ , \label{fano}
\end{eqnarray}
with $\eta$ being the asymmetry parameter.

\begin{figure}[htb!]
\includegraphics[angle=0, width=\linewidth]{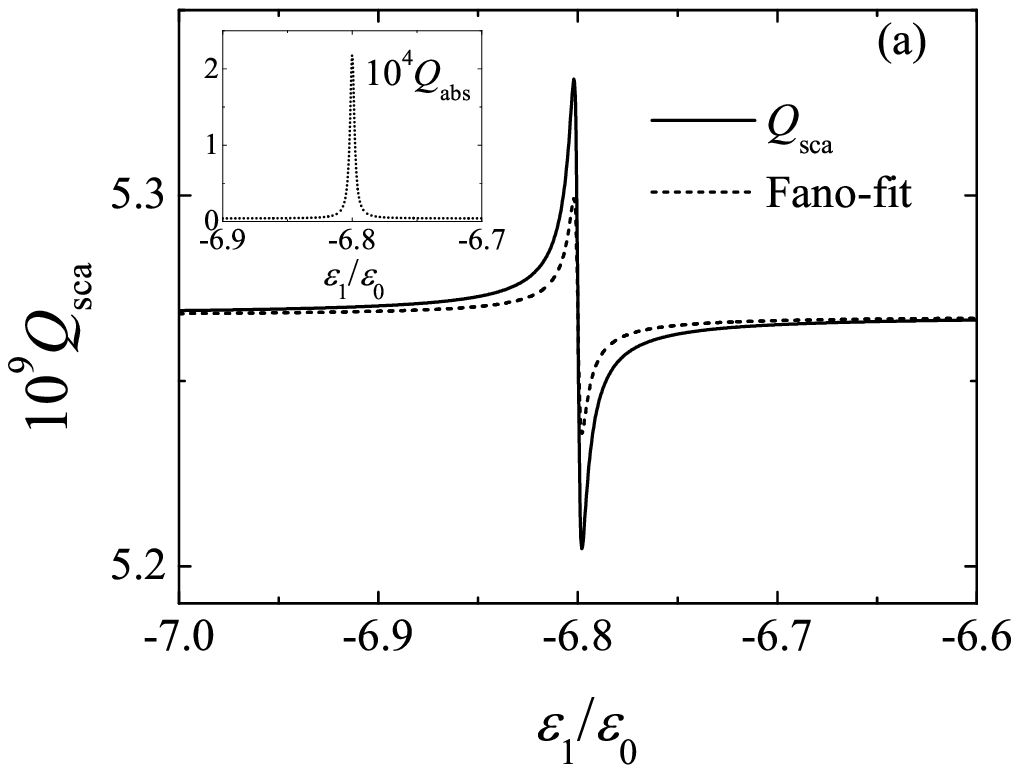}\vfill
\includegraphics[angle=0, width=\linewidth]{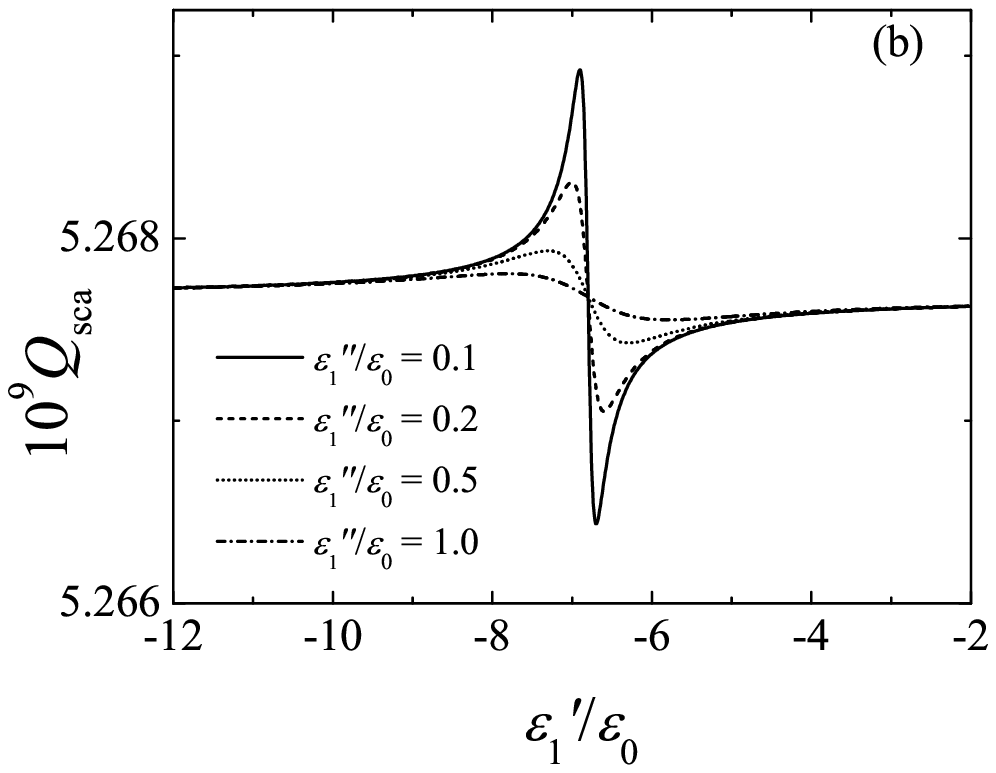}
\caption{Scattering efficiency $Q_{\rm sca}$ in the vicinities of the electric dipole resonance for a plasmonic sphere $(\epsilon_1,\mu_1=\mu_0)$ coated with a lossy dielectric shell $(\epsilon_2/\epsilon_0=3.4+0.001\imath;\mu_2/\mu_0=1)$ as a function of $\epsilon_1$ in the the small-particle limit (size parameters $ka=10^{-4}$, $kb=10^{-2}$).
(a) The dipole resonance for a lossless core $(\epsilon_1=\epsilon_1')$ is well described by the Fano lineshape,  Eq.~(\ref{fano}) (dotted line).
The inset shows the absorption efficiency as a function of $\epsilon_1$.
(b) $Q_{\rm sca}$ as a function of the real part of the core electric permittivity ($\epsilon_1'$) for different values of its imaginary part ($\epsilon_1''$) in the vicinities of the electric dipole resonances.} \label{fig2}
\end{figure}

In Fig.~\ref{fig2}(a), the scattering efficiency $Q_{\rm sca}$ for a coated sphere in the Rayleigh limit ($y\ll1$, $|m_1|x\ll1$ and $|m_2|y\ll1$) is plotted as a function of the electric permittivity of the core ($\epsilon_1,\mu_1=\mu_0$) in the vicinity of the dipole resonance.
This plot confirms that $|a_1|^2$ exhibits an unconventional Fano resonance, which is well described by the Fano lineshape, Eq.~(\ref{fano}).
Also, the plot in Fig.~\ref{fig2}(a) demonstrates that, for coated spheres without spatial dispersion, the existence of unconventional Fano resonances does not necessarily require large values of the dielectric permittivities, a situation that is typically difficult to achieve in natural media.
This result is in contrast to the case of homogeneous spheres, where such resonances are predicted to occur only beyond the Rayleigh limit~\cite{tribelsky}.
Indeed, the plot in Fig.~\ref{fig2}(a) shows that unconventional Fano resonances in light scattering by coated spheres can exist even in the Rayleigh limit ($y\ll1$, $|m_1|x\ll1$ and $|m_2|y\ll1$), corroborating the previous analytical result.

The effect of finite dissipation is investigated in Fig.~\ref{fig2}(b), where $Q_{\rm sca}$ is shown as a function of the real part of the core electric permittivity ($\epsilon_1'$), for different values of its imaginary part ($\epsilon_1''$), in the vicinities of the dipole resonances that result in the Fano lineshape in $|a_1|^2$.
From this plot, it is clear that these resonances are quite robust against losses so that they are not completely washed out in the presence of finite dissipation.

For homogeneous spheres, unconventional Fano resonances in $|a_1|^2$ are expected to manifest themselves not only in the the scattering efficiency $Q_{\rm sca}$, but also in the extinction efficiency $Q_{\rm ext}$ even in the presence of finite dissipation~\cite{tribelsky}.
For coated spheres in the Rayleigh limit with $S\ll1$, the plot in Fig.~\ref{fig2}(a) reveals that extinction is dominated by absorption, {\it i.e.} typically $Q_{\rm sca} \ll Q_{\rm abs}$, so that $Q_{\rm ext} \approx Q_{\rm abs}$, even in the weakly absorption approximation.
We have numerically verified that $Q_{\rm sca}\approx (kb)^3Q_{\rm abs}$ in off-resonant regions.
A similar effect is well-known for sufficiently small homogeneous spheres of radius $R$, where $Q_{\rm abs}\propto kR$ and $Q_{\rm sca}\propto (kR)^4$~\cite{bohren}.
As a result, unconventional Fano resonances in $Q_{\rm sca}$ tend to be unnoticed in $Q_{\rm ext}$, which, like $Q_{\rm abs}$, exhibits a Lorentzian profile in the vicinities of the dipole resonance in $|a_1|^2$.
Hence, the observation of unconventional Fano resonances in extinction for weakly absorbing coated spheres with $S\ll1$ requires that $Q_{\rm ext} \approx Q_{\rm sca}$.
This condition can be achieved imposing that $ka\ll1$ for the core and that $kb$ is large enough to only amplify the dipole resonance, but small enough to ensure the validity of the first order approximation.

We emphasize that the unconventional Fano resonances in coated spheres are not a result of interferences between plasmon modes in the different layers of the system (core and shell), as it is generally expected for multilayered particles, such as nanoshells~\cite{luk}.
In contrast, here the Fano-like resonance in $Q_{\rm sca}$ in the Rayleigh limit [Fig.~\ref{fig2}(a)] occurs due to self-interferences between partial waves within the shell.
Mathematically, they arise from partial waves generated from both Bessel and Neumann special functions.
Within the core, wave interferences are generated only from Bessel functions.
The existence of these wave interferences within the shell, that eventually lead to Fano-like resonances in $Q_{\rm sca}$, is confirmed by the analytical expressions for the angle-averaged field intensities inside the core and shell, Eqs.~(\ref{E1-med}) and (\ref{E2-med}), respectively.
Indeed, Eq.~(\ref{E2-med}) explicitly encodes interferences between internal field coefficients, which are associated with the scattering ones by the boundary condition
\begin{eqnarray}
   a_n=\frac{\psi_n(y)}{\xi_n(y)}-\frac{\left[\psi_n(m_2y)-A_n\chi_n(m_2y)\right]}{\mu_2\xi_n(y)}g_n\ .\label{an-fano}
\end{eqnarray}
The corresponding condition for $b_n$ is obtained by replacing $(a_n,A_n)$ with $(b_n,B_n)$ and $(\mu_2,g_n)$ with $(m_2,f_n)$.
Inspecting Eq.~(\ref{an-fano}), one can see that $|a_n|^2$ contains interference terms between the internal coefficients $g_n$ and $w_n=A_ng_n$ of the shell.
The presence of these interference terms, such as {\it e.g.} $g_1^*w_1=A_1|g_1|^2$, shows that an off-resonant $A_1$ may play the role of a broad resonance that interferes with a sharp one in $g_1$.
Since the denominators of $a_1$ and $g_1$ are the same [see Eqs.~(\ref{an}) and (\ref{gn})], an unconventional Fano resonance in $|a_1|^2$ may occur for coated spheres even in the Rayleigh limit.
Indeed, for a dielectric shell and dispersive core in the Rayleigh limit, one has $|\widetilde{\gamma}/\epsilon_0|\ll1$ if $S\ll1$; in this case, $A_1$ plays the role of the broad resonance in the Fano effect.

\section{Connection to off-resonance field enhancement}
\label{off-res}

\begin{figure}[htb!]
\includegraphics[angle=0, width=\linewidth]{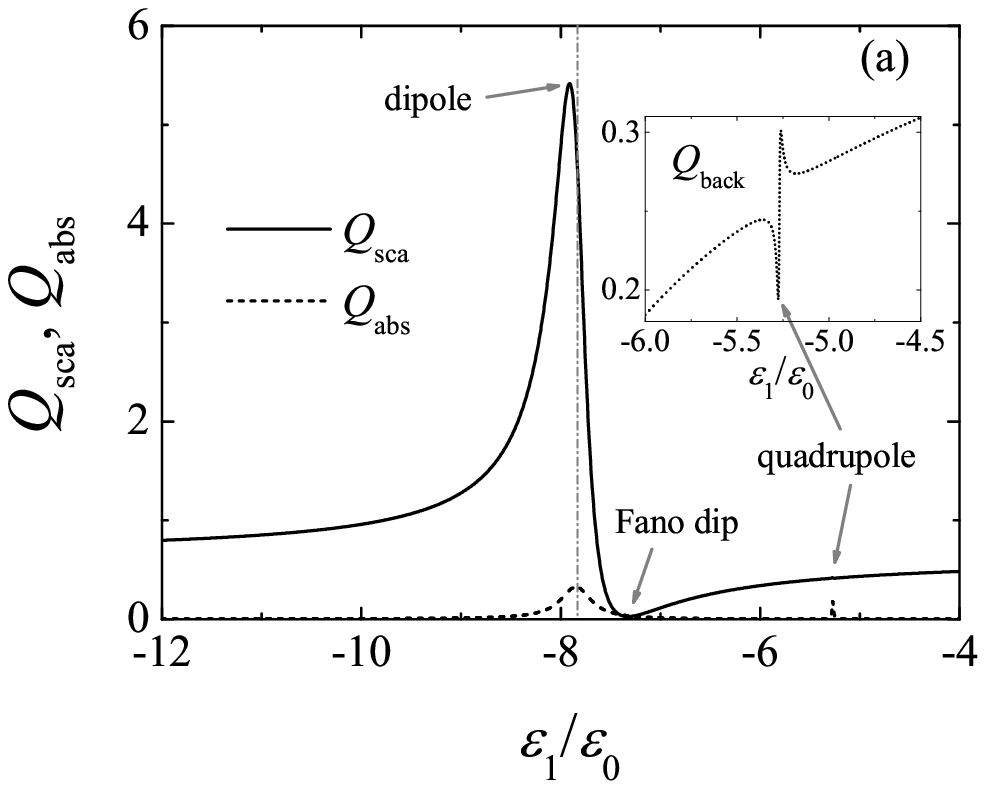}\vfill
\includegraphics[angle=0, width=\linewidth]{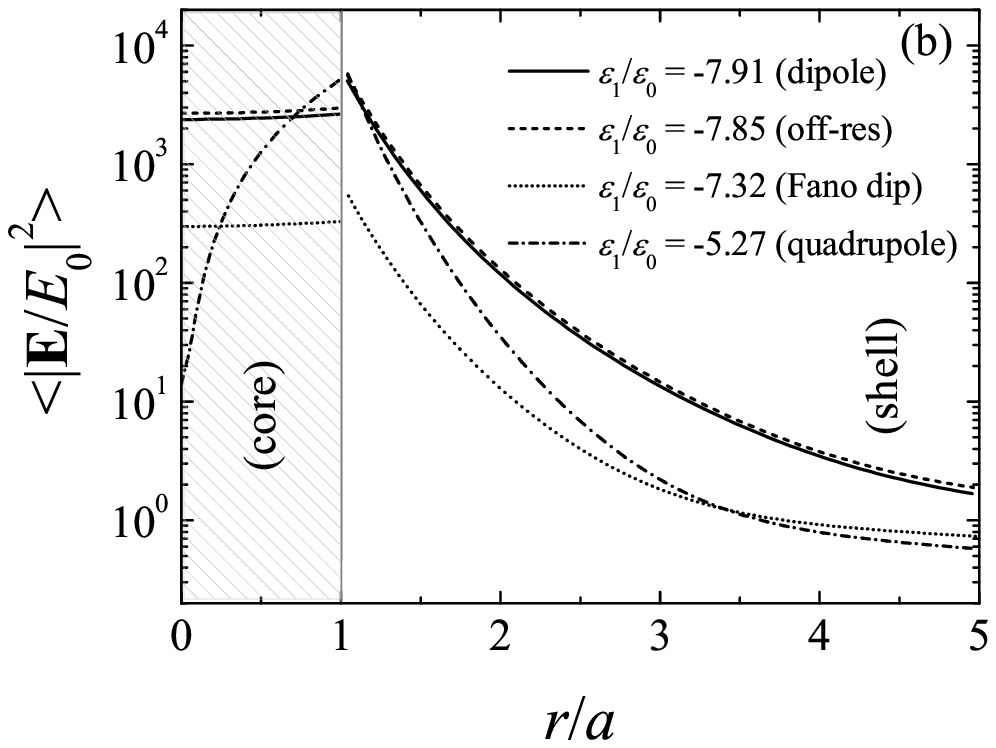}
\caption{Fano-like resonances and off-resonance field enhancement for a metallic non-absorbing sphere $(\epsilon_1/\epsilon_0<0;\mu_1=\mu_0)$ coated with a lossy dielectric shell $(\epsilon_2/\epsilon_0=3.4+0.004\imath;\mu_2/\mu_0=1)$ with size parameters $ka=0.2$ and $kb=1$.
(a) $Q_{\rm sca}$ and $Q_{\rm abs}$ as a function of $\epsilon_1$.
The dotted vertical line indicates the position of the unconventional Fano resonance.
The inset shows the backscattering efficiency $Q_{\rm back}$ as a function of $\epsilon_1$ at the vicinities of the quadrupole resonance.
(b) Angle-averaged electric field intensity as a function of the distance to the center of the coated sphere for $\epsilon_1/\epsilon_0=-7.91$ (sharp dipole resonance), $\epsilon_1/\epsilon_0=-7.85$ (unconventional Fano resonance), $\epsilon_1/\epsilon_0=-7.32$ (Fano dip), and $\epsilon_1/\epsilon_0=-5.27$ (conventional Fano resonance).} \label{fig3}
\end{figure}

For coated spheres consisting of metal-dielectric composites, it was demonstrated that conventional Fano resonances are at the origin of an off-resonant field enhancement.
Due to the Fano effect, the maximal field enhancement inside the scatterer does not necessarily corresponds to a resonance in the extinction efficiency, contrary to the common belief~\cite{miroshnichenko}.
In the plot in Fig.~\ref{fig3}(a), we investigate whether unconventional Fano resonances can also induce a similar effect.
Since $Q_{\rm ext}=Q_{\rm sca}+Q_{\rm abs}$, only the efficiencies $Q_{\rm sca}$ and $Q_{\rm abs}$ are calculated for a non-absorbing metallic sphere $(\epsilon_1<0;\mu_1=\mu_0)$ coated with a dielectric shell $(\epsilon_2=3.4+0.004\imath;\mu_2=\mu_0)$ as a function of $\epsilon_1$.
The asymmetric lineshape of $Q_{\rm sca}$ in the vicinities of the electric dipole resonance is a result of the unconventional Fano effect, as it results from the interference of two dipole resonances ({\it i.e.} a resonance in $|a_1|^2$).
The unconventional Fano resonance frequency occurs for $\epsilon_1/\epsilon_0=-7.85$, a value in between the ones corresponding to maximal and minimal scattering (constructive and destructive interferences): $\epsilon_1/\epsilon_0=-7.91$ (sharp dipole resonance) and $\epsilon_1/\epsilon_0=-7.32$ (broad dipole resonance), respectively.
For simplicity, we refer to the minimum scattering in the Fano curve as the Fano dip.
At the same time, the value $\epsilon_1/\epsilon_0=-7.85$ corresponds to a peak in $Q_{\rm abs}$ which, by Eq.~(\ref{Qabs-exact}), is related to maximal field enhancement inside the system.
This result demonstrates that unconventional Fano resonances can also lead to off-resonance field enhancement.
This conclusion is corroborated by the analysis of the plot in Fig.~\ref{fig3}(b), which shows the angle-averaged electric field as a function of the distance to the center of the coated sphere for different values of $\epsilon_1$.
The maximal field enhancement does not correspond to maximal scattering, but rather it is achieved for $\epsilon_1/\epsilon_0=-7.85$, precisely the position of the unconventional Fano resonance and the peak in $Q_{\rm abs}$.

The plot in Fig.~\ref{fig3}(a) also reveals the existence of a conventional Fano resonance at $\epsilon_1/\epsilon_0=-5.27$.
It appears in the backscattering efficiency $Q_{\rm back}$ [see inset of Fig.~\ref{fig3}(a)] and results from the interference between a broad dipole resonance with a sharp quadrupole one.
This conventional Fano resonance not only appears in $Q_{\rm back}$, as expected from Ref.~\cite{miroshnichenko}, but also is indicated as a peak in $Q_{\rm sca}$, further confirming the analysis of Sec.~\ref{fano-sec}.
From Eq.~(\ref{Qabs-exact}) and since it does not appear as a peak in $Q_{\rm sca}$, this conventional Fano resonance is related to an off-resonance field enhancement inside the scatterer.
For these parameters, one can see from the plot in Fig.~\ref{fig3}(b) that the off-resonance enhancement of the electric field related to conventional Fano resonances is approximately one order of magnitude smaller than the one achieved with the unconventional Fano effect.
One can partially explain this result by the fact that the unconventional Fano effect involves two dipole resonances while the conventional one involves dipole and quadrupole (which is of higher order) resonances.
However, the amount of field stored at the quadrupole resonance depends on the absorption in the core $(\epsilon_1;\mu_1=\mu_0)$ and shell $(\epsilon_2/\epsilon_0=3.4+0.004\imath;\mu_2=\mu_0)$.
If $\epsilon_2''<0.004$ in our example of lossless core ($\epsilon_1''=0$), one may obtain at the quadrupole resonance an electromagnetic energy stored inside the particle even greater than the one at the dipole resonance.
In this non-realistic case, nevertheless, the quadrupole resonance would appear as a peak in $Q_{\rm sca}$.
This indicates that, for a lossless dielectric shell, a finite absorption in the dispersive core is necessary to minimize the scattering at the quadrupole resonance in small-particle limit.
Also, for a metallic core and dielectric shell, an off-resonance field enhancement near the quadrupole resonance emerges only for $S\ll1$~\cite{miroshnichenko}, which is the same condition we have obtained for unconventional Fano resonances in coated spheres.
This leads to the possibility of combining these two interference effects.

Although Fano-like resonances in $Q_{\rm back}$ exist even for homogeneous spheres~\cite{luk}, ``hidden'' quadrupole resonances in $Q_{\rm sca}$ (they appear only in $Q_{\rm abs}$ and leads to an off-resonance field enhancement) are usually achieved in layered particles, as we show for coated spheres in the following.
In the Aden-Kerker solution, for a given $kb$ in off-resonant regions, $ka\ll1$ leads to $|A_n|\ll1$ and $|B_n|\ll1$, where $A_n$ and $B_n$ are defined in Eqs.~(\ref{An}) and (\ref{Bn}).
Omitting the leading order $\mathcal{O}(x^7)$ in Rayleigh limit, we obtain the approximations:
\begin{eqnarray}
A_1&\approx&{2(m_2x)^3}\left(\frac{\epsilon_1-\epsilon_2}{\epsilon_1+2\epsilon_2}\right)\left[\frac{1}{3}+\frac{(m_2x)^2}{5}\left(\frac{\epsilon_1-2\epsilon_2}{\epsilon_1+2\epsilon_2}\right)\right]\ ,\label{A1-ap}\\
A_2&\approx&-\frac{(m_2x)^5}{15}\left(\frac{\epsilon_1-\epsilon_2}{2\epsilon_1+3\epsilon_2}\right)\ .\label{A2-ap}
\end{eqnarray}
Resonances in $A_1$ and $A_2$ are obtained when the real parts of their denominators vanish: $\epsilon_1'(\omega)=-2\epsilon_2'$ and $\epsilon_1'(\omega)=-3\epsilon_2'/2$, respectively.
If we have no absorption ($\epsilon_1''=\epsilon_2''=0$) for finite $x=ka$, both $A_1$ and $A_2$ diverge at the resonance and there is no ``hidden'' quadrupole resonance in $Q_{\rm sca}$.
Instead, a sharp resonance is observed in $|a_2|^2$.
Consider now a weakly absorbing media such that the first term in Eq.~(\ref{A1-ap}) does not depend on $ka$ at the dipole resonance: $x^3\approx(\epsilon_1''+2\epsilon_2'')$.
This leads to $x^5\ll(2\epsilon_1''+3\epsilon_2'')$ and Eq.~(\ref{A2-ap}) is partially suppressed at the quadrupole resonance, whereas the second term inside brackets in Eq.~(\ref{A1-ap}) yields $A_1\sim1/x$ in off-resonant regions (broad dipole resonance).
Therefore, in this approximation, the dipole resonance is huge compared to the quadrupole one in $Q_{\rm sca}$ and one may achieve interference between $a_1$ and $a_2$ in $Q_{\rm back}$.
This result provides a simple condition to obtain the so-called ``dark states'' in the small-particle limit.
Indeed, in the plot in Figs.~\ref{fig3}, we have assumed $x^3=(\epsilon_1''+2\epsilon_2'')=8\times10^{-3}$, and similar conditions can always be fulfilled for dispersive cores.

\section{Optimal field enhancement with negative refractive index materials}
\label{negative}

\begin{figure}[htb!]
\includegraphics[angle=0, width=\linewidth]{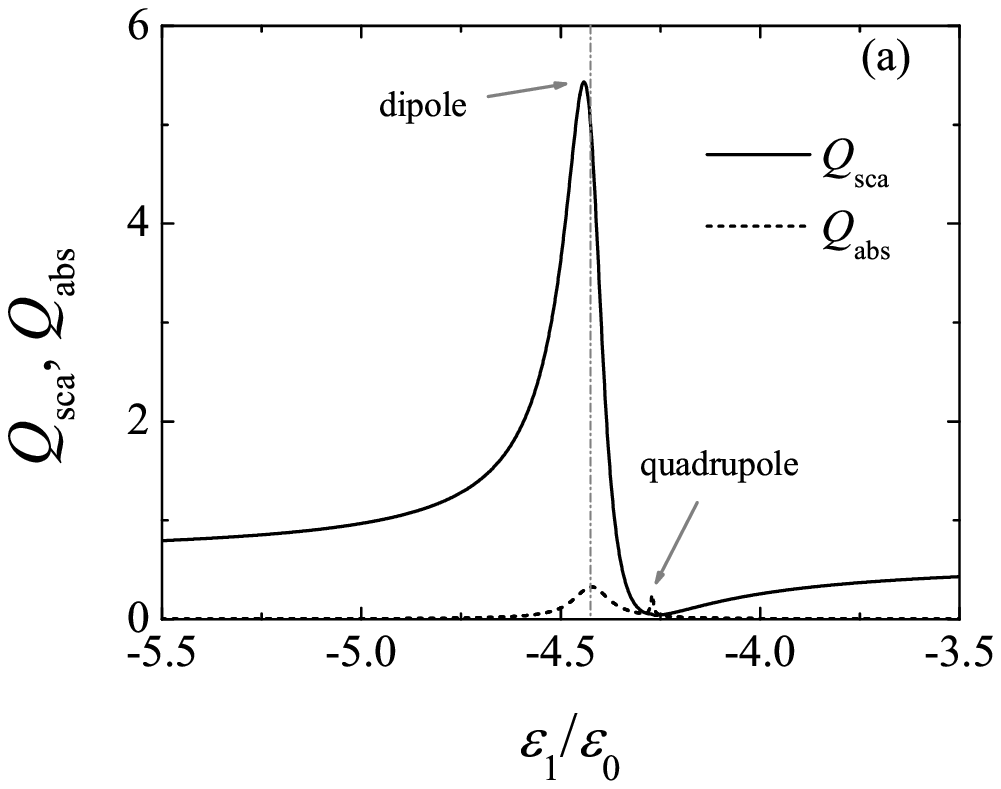}\vfill
\includegraphics[angle=0, width=\linewidth]{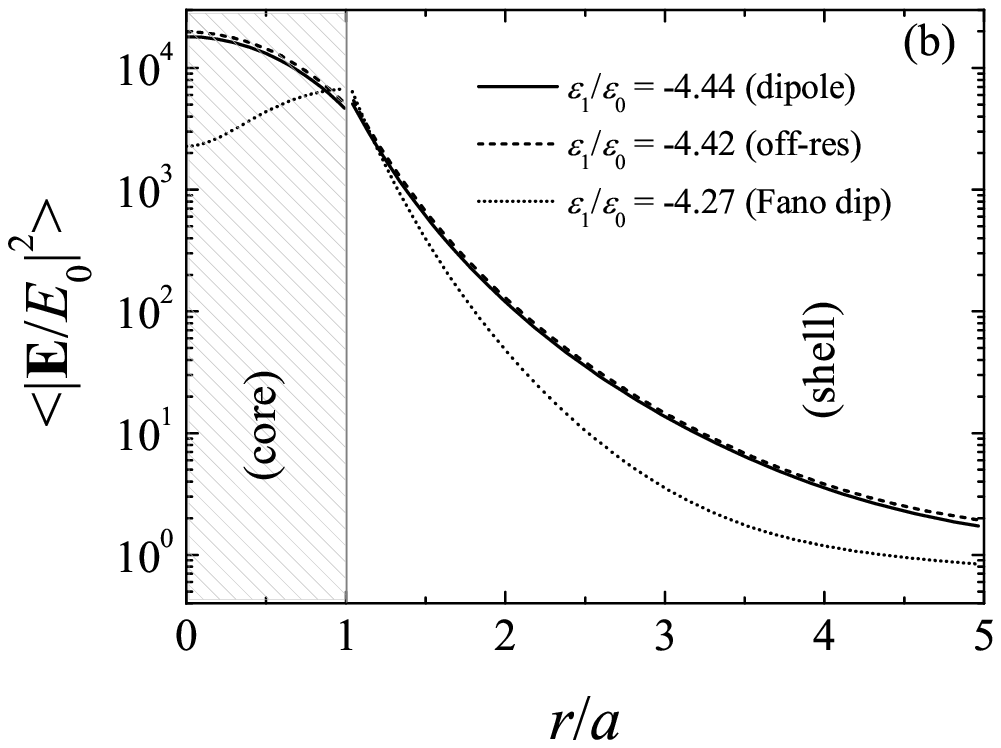}
\caption{Fano-like resonances and off-resonance field enhancement for non-absorbing sphere with negative refractive index $(\epsilon_1/\epsilon_0<0;\mu_1/\mu_0=-21)$ coated with a lossy dielectric shell $(\epsilon_2/\epsilon_0=3.4+0.004\imath;\mu_2/\mu_0=1)$ with size parameters $ka=0.2$ and $kb=1$.
(a) $Q_{\rm sca}$ and $Q_{\rm abs}$ as a function of $\epsilon_1$.
The dotted vertical line indicates the position of the unconventional Fano resonance.
(b) Angle-averaged electric field intensity as a function of the radial distance for $\epsilon_1/\epsilon_0=-4.44$ (sharp dipole resonance), $\epsilon_1/\epsilon_0=-4.42$ (unconventional Fano resonance), and $\epsilon_1/\epsilon_0=-4.27$ (Fano dip).} \label{fig4}
\end{figure}

Many applications in photonics depends crucially on their capacity of enhancing light confinement in small volumes with low losses.
In the previous section, we have demonstrated that unconventional Fano resonances lead to a large off-resonance enhancement of the electromagnetic field intensity, even larger than the one achieved for conventional Fano resonances.
To achieve an optimal field enhancement inside the coated sphere, let us investigate the conditions for simultaneous occurrence of conventional and unconventional Fano resonances at the same frequency.
With this aim, we consider a plasmonic dispersive core $[\epsilon_1(\omega),\mu_1(\omega)]$ coated with a nondispersive dielectric shell ($\epsilon_2,\mu_2=\mu_0$) and impose that a dipolar antiresonance in $a_1$ coincides with a quadrupolar resonance in $a_2$ at a given frequency $\bar{\omega}$.
The antiresonance in $|a_1|^2$ (Fano dip) for $ka\ll1$ ({\it i.e.} $S\ll1$) and $|m_2|\approx1$ approximately coincides with the antiresonance in $A_1$ (that plays the role of a broad resonance in the unconventional Fano effect): $\widetilde{m}_2D_1(m_1x)=\widetilde{m}_1D_1(m_2x)$.
From this expression, one can see that in nonmagnetic media ($\widetilde{m}_q=m_q$) the condition $A_1=0$ can only be fulfilled for $m_1=m_2$, {\it i.e.} for homogeneous spheres.
Since homogeneous spheres do not exhibit off-resonance field enhancement due to the conventional Fano effect~\cite{miroshnichenko}, for the purpose of achieving maximal field enhancement one should consider $\mu_1\not=\mu_2$.
For the sake of simplicity, in the following we consider the approximation $D_1(\rho)\approx(2/\rho)-(\rho/5)$~\cite{bohren}.
Within this approximation, the condition for the existence of an antiresonance in $a_1$ is $x^2\approx 10\epsilon_0\mu_0(\epsilon_2'-\epsilon_1')/[\epsilon_1'\epsilon_2'(\mu_1'-\mu_2')]>0$.
Since the core is magnetic ($\mu_1\not=\mu_{0}$) we also impose that $b_1\to0$, which is the condition for the emergence of a broad resonance in $b_1$; this yields $x^2\approx 10\epsilon_0\mu_0(\mu_2'-\mu_1')/[\mu_1'\mu_2'(\epsilon_1'-\epsilon_2')]>0$.
Altogether, both antiresonance conditions for $A_1$ ($a_1$) and $B_1$ ($b_1$) are equivalent to $(\epsilon_1'\mu_1')(\epsilon_2'\mu_2')>0$.
For a dielectric shell ($\epsilon_2'>\epsilon_0>0$) with thickness parameter $0<S<1$, the second Fröhlich mode occurs in the interval $-3\epsilon_2'/2<\epsilon_1'(\omega_{\rm F})<-3\epsilon_0/2$~\cite{bohren}.
Since $\epsilon_2'\mu_2'>0$ and $\epsilon_1'<0$, it follows from this conditions that $\mu_1'<0$.
As a result, we conclude that for $ka\ll1$ and a dielectric shell $(m_2'>0)$ a sharp plasmon resonance in $a_2$ will only coincide with an antiresonance in $a_1$ provided the dispersive core has a negative refractive index ($m_1'<0$).

Figure~\ref{fig4}(a) shows the plots of the scattering $Q_{sca}$ and absorption $Q_{abs}$ efficiencies for non-absorbing sphere with negative refraction index $(\epsilon_1/\epsilon_0<0;\mu_1/\mu_0=-21)$ coated with a lossy dielectric shell $(\epsilon_2=3.4+0.004\imath;\mu_2=\mu_0)$ as a function of $\epsilon_1$.
We demonstrate that the dipole and quadrupole resonance can be brought together provided the core has a negative index, confirming the previous analytical result.
Hence conventional and unconventional Fano resonances can occur at the same value of $\epsilon_1$.
As a consequence, there is a huge field enhancement near these resonances, as confirmed by the analysis of the plot in Fig.~\ref{fig4}(b), and one may achieve the maximum field stored inside the particle with minimum scattering (Fano dip in $\epsilon_1/\epsilon_0=-4.27$).

\section{Conclusions}
\label{conclusions}

In this paper we have investigated Fano-like resonances in light scattering by coated spheres composed of a dispersive plasmonic core and a dielectric shell.
Using the Aden-Kerker solution, we have derived an analytical expression for the absorption efficiency $Q_{\rm abs}$ as a function of the internal fields.
This expression explicitly contains interference terms between internal Aden-kerker coefficients of the shell that are not washed out by the average among all possible directions and polarizations.
This result shows that Fano-like resonances, which result from interference between electromagnetic modes inside the scatterer, can be identified in the total cross-sections, which contrasts to the common belief that field interferences, and hence Fano resonances, can only be identified in differential scattering spectra ({\it e.g.} the radar backscattering efficiency).
Specially, we have demonstrated that unconventional Fano resonances, recently discovered for homogeneous spheres~\cite{tribelsky}, can also occur for coated spheres.
These resonances arise from the interference between two electromagnetic modes with the same multipole moment within the shell and not from interactions between various plasmon modes of different layers of the particle.
In contrast to the case of homogeneous spheres, the existence of unconventional Fano resonances for coated spheres do not require large electric permittivities so that they can occur even in the Rayleigh limit.
As for conventional Fano resonances~\cite{miroshnichenko}, unconventional Fano resonances in coated spheres can induce an off-resonance field enhancement.
This enhancement is, nevertheless, approximately one order of magnitude larger than the one achieved with conventional Fano resonances for lossy scatterers, since unconventional Fano resonances involves multipole moments with the same order.
Finally, we have examined the conditions for an optimal field enhancement inside the scatterer.
We show that unconventional and conventional Fano resonances can occur simultaneously provided the core has a negative refraction index, leading to a maximal field enhancement.
We believe that our results could be relevant not only for a better understanding of Fano resonances in light scattering but also to the development of novel applications in photonics.

\section*{Acknowledgments}

The authors thank the fruitful discussions with M. H. Y. Moussa and acknowledge the Brazilian agencies for support.
TJA holds grants from Fundação de Amparo à Pesquisa do Estado de São Paulo (FAPESP) (2010/10052-0), ASM from Conselho Nacional de Desenvolvimento Científico e Tecnológico (CNPq) (305738/2010-0), and FAP from Fundação de Amparo à Pesquisa do Estado do Rio de Janeiro (FAPERJ) (E-26/111.463/2011).

\end{document}